\documentclass[12pt]{article}
\usepackage{amsmath,epsfig,graphicx,amssymb}

\newcommand{\beq}{\begin{equation}}
\newcommand{\eeq}{\end{equation}}
\newcommand{\ben}{\begin{eqnarray}}
\newcommand{\een}{\end{eqnarray}}
\newcommand{\bes}{\begin{subequations}}
\newcommand{\ees}{\end{subequations}}
\newcommand{\bFig}{\begin{figure}}
\newcommand{\eFig}{\end{figure}}

\date{}
\begin{document}

\title{Entangling successive single-photons from quantum dots}
\author{Partha Ghose\footnote{partha.ghose@gmail.com}  \\
Centre for Astroparticle Physics and Space Science (CAPSS),\\Bose Institute, \\ Block EN, Sector V, Salt Lake, Kolkata 700 091, India.}
\maketitle
\begin{abstract}
A method is proposed for generating and discriminating Bell states of high fidelity from consecutive single-photons generated in a semiconductor quantum dot. The use of a non-symmetric beam splitter is found to be essential and sufficient, and no nonlinear optical process such as SPDC is required at any stage. This should be of considerable importance for testing the foundations of quantum mechanics. 
\end{abstract}

\section{Introduction}

Entanglement generation has become extremely important as a resource for testing the foundations of quantum theory and for quantum information processing, specially for quantum communication and quantum cryptography \cite{qc1,qc2,qc3}. In most applications strongly correlated and entangled photon pairs are generated by spontaneous parametric downconversion (SPDC) of a laser pump beam in a nonlinear crystal \cite{pdc1, pdc2,pdc3,pdc4,pdc5,pdc6,pdc7,pdc8}. Other methods that have been used are two-photon emission from atoms \cite{2photon1,2photon2} and entangled photon pair generation in fibres \cite{fan, chen}. Of special interest in this context is the rapid generation of {\em single-photons} from semiconductor quantum dots in a microcavity with the feature that consecutive photons have the same wavepacket and are largely indistinguishable \cite{qdots}. Here we propose a method of generating Bell states of high fidelity from a pair of such consecutively generated single-photons by making them incident on a {\em non-symmetric} and lossless beam splitter from {\em opposite} sides at the same time, and then using post-selection. For convenience and to establish the notation, the required input-output relations, originally derived by Zeilinger \cite{zeilinger} for general lossless beam splitters, are rederived for this case. This method does not involve any nonlinear optical process like SPDC, and should be of considerable interest for linear optical computing and testing the foundations of quantum mechanics.   

That entangled states can be generated from a direct product state was first shown by Yurke and Stoler \cite{yurke}. Their method requires, however, the use of already entangled pairs and entanglement swapping through a Bell basis measurement. In the method proposed here, the initial states are unentangled single-photons, and non-maximal entanglement is fist generated through a {\em non-symmetric} beam splitter without requiring any nonlinear process. Bell states of high fidelity are then discriminated by post-selection. The utility of non-symmetric beam splitters for linear optical quantum computation has been noted earlier also \cite{knill}, though not for generating entanglement and Bell states.
\section{The Method}
\begin{figure}
\centering
{\includegraphics[scale=0.6]{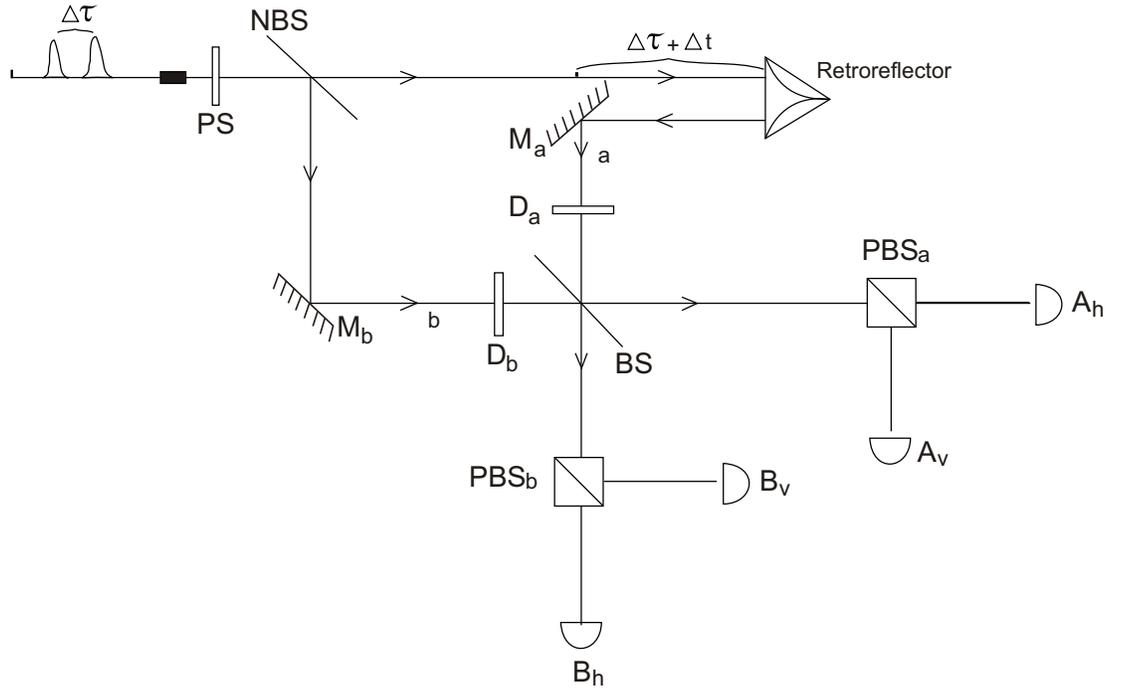}}
\caption{\label{Figure}{\footnotesize Schematic diagram of the experiment: Two consecutive coherent photons separated by a small time interval $\Delta \tau$ pass through a polarization selector $PS$, a non-polarizing beam splitter $NBS$ and a system of reflectors and two devices $D_a$ and $D_b$ such that two arbitrary polarization states are incident on a lossless, nonpolarizing and non-symmetric beam splitter $BS$ from opposite sides $a$ and $b$ at the same time. The reflected and transmitted states are sent through 50-50 polarizing beam splitters $PBS_a$ and $PBS_b$ and finally detected by the single-photon detectors $A_h, A_v, B_h$ and $B_v$.}}
\end{figure}

Let a pair of consecutive collinear single-photon pulses from a quantum dot, separated by a small time interval $\Delta \tau$, arrive through a single-mode fibre and a polarization selector $PS$ at a non-polarizing $50$-$50$ beam-splitter $NBS$, as shown in Figure 1. Let the pulses overlap and interfere on a second non-symmetric, lossless and nonpolarizing beam-splitter $BS$ from opposite sides (labelled by $a$ and $b$) after (i) reflection from two lossless mirrors $M_a$ and $M_b$ and (ii) after passing through devices $(D_a, D_b)$ which can rotate the polarizations and prepare arbitrary polarization states in each path. The overlap can be ensured by adjusting the optical path difference $\Delta \tau + \Delta t$ between the two arms, as in a Michelson interferometer. The following notation will be used. Since the Hilbert space of a single-photon is the tensor product ${\cal{H}}_{path}\otimes {\cal{H}}_{pol}$ of its path and polarization Hilbert spaces, we will write $\vert a\rangle^P_{in} \equiv \vert a\rangle_{in}\otimes \vert P\rangle,\,\vert b\rangle^P_{in} \equiv \vert b\rangle_{in}\otimes \vert P\rangle$ where $P$ stands for $H$ or $V$, the polarization bases, and $\vert a\rangle_{in}$ and $\vert b\rangle_{in}$ are the incoming spatial modes corresponding to the two paths $a = \langle {\bf x}\vert a\rangle$ and $b = \langle {\bf x}\vert b\rangle$. Then the incoming state at the ports $a$ and $b$ of $BS$ can be written as a tensor product state
\ben
\vert \Psi\rangle_{in} &=&
(\alpha\vert a\rangle_{in}^V + \beta \vert a \rangle_{in}^H)\otimes (\alpha^\prime\vert b\rangle_{in}^V + \beta^\prime \vert b\rangle_{in}^H)\nonumber\\
&=& \alpha\alpha^\prime\vert a\rangle_{in}^V\otimes \vert b\rangle_{in}^V + \beta\beta^\prime \vert a\rangle_{in}^H\otimes\vert b\rangle_{in}^H\nonumber\\ &+& \alpha\beta^\prime
\vert a\rangle_{in}^V\otimes\vert b\rangle_{in}^H + \beta\alpha^\prime \vert a\rangle_{in}^H\otimes\vert b\rangle_{in}^V,
\een 
where ($\alpha,\beta, \alpha^\prime, \beta^\prime$) are tunable complex amplitudes with $\vert\alpha\vert^2 + \vert\beta\vert^2 = 1$, $\vert\alpha^\prime\vert^2 + \vert\beta^\prime\vert^2 = 1$. In quantum theory the incoming states are converted to the ougoing states by a unitary operator. Following Zeilinger \cite{zeilinger} (but changing his notation to suit our purpose), we let the incident states be piloted to the outgoing states by the general matrix beam splitter relation

\[\left(\begin{array}{c}
 a\\b 
\end{array} \right)_{out}^{P} = \left(\begin{array}{cc}
r^a_{p}\,\,\,\,\,\,\,\,\,\,\,\,t^b_{p} \\t^a_{p} \,\,\,\,\,\,\,\,\,\,\,\,\, r^b_{p}
\end{array} \right)\left(\begin{array}{c}
 a\\b 
\end{array} \right)_{in}^{P}\equiv B_{p} \left(\begin{array}{c}
 a\\b\ 
\end{array} \right)_{in}^{P}\]
where $r^a_p, r^b_p, t^a_p, t^b_p$ are the reflection and transmission coefficients of the polarization state $P$ on the two sides of the beam splitter. The underlying dynamics is defined by $B_p = {\rm exp \{i\epsilon H_p \}}$ with $H_p^\dagger = H_p$ as the Hamiltonian. Inverting this, we get

$$B_p^{-1} = \Delta_p^{-1}\left(\begin{array}{cc}
 r^b_p\,\,\,\,\,\,-t^b_p \\-t^a_p \,\,\,\,\,\,\,\,\,\,\,r^a_p
\end{array} \right) = \left(\begin{array}{cc}
 r^{a*}_p\,\,\,\,\,\,\,\,t^{a*}_p \\t^{b*}_p \,\,\,\,\,\,\,\,\,\,\,r^{b*}_p
\end{array}\right)$$ 
with $\Delta_p = r^a_p r_p^b - t^a_p t^b_p = e^{i\chi}$ because the determinant of a unitary matrix is of modulus unity. Since $\chi$ is an overall phase, we can set $\chi =0$ so that 
\beq
\Delta_p = 1 \label{norm}. 
\eeq
Unitarity of $B_p$ implies
\beq
r_p^b = r^{a*}_p,\,\,\,\, t_p^a = - t_p^{b*}.\label{u1}
\eeq 
One therefore has, as first shown by Zeilinger,
\ben
\vert r^a_p\vert = \vert r^b_p\vert,\,\,\,\,\vert t^a_p\vert &=& \vert t^b_p\vert,\label{u2}\\
\delta_{t^a_p} -\delta_{r^a_p} + \delta_{t^b_p} -\delta_{r^b_p} &=& \pi, \label{u3}
\een
where $\delta$'s are phases. Apart from this relation, any phase relation between the transmitted and reflected states is allowed. Accordingly, we will set all the reflection amplitudes to be real and $t^{a,b}_p = -i\vert t^{a,b}_p\vert$. Henceforth, therefore, to simplify the notation, we will omit the superscripts $a$ and $b$ and also write $r_p, t_p$ etc instead of $\vert r_p\vert, \vert t_p\vert$. Thus, 

$$B_p^{-1} = \Delta_p^{-1}\left(\begin{array}{cc}
 r_p\,\,\,\,\,\,\,\,\,\,it_p \\it_p \,\,\,\,\,\,\,\,\,\,\,r_p
\end{array} \right)$$
with $\Delta_p = r_p^2 + t_p^2 = 1$. For a 50-50 beam splitter, $r_p = t_p = \frac{1}{\sqrt{2}}$ and
\beq
\delta_{t^a_p} -\delta_{r^a_p } = \pi/2,\label{sm1}
\eeq
which is widely used in the literature. {\em Violation of this set of relations by a non-symmetric beam splitter will turn out to be of crucial importance for our purpose}. 

We have, therefore, the following relations for a non-symmetric beam splitter:
\ben
\vert a\rangle^p_{in} &=& r_p\vert a\rangle^p_{out} + it_p\vert b\rangle^p_{out},\\
\vert b\rangle^p_{in} &=& i t_p\vert a\rangle^p_{out} + r_p\vert b\rangle^p_{out}.
\een
Finally, omitting the subscript $out$ from all kets and the symbol $\otimes$ between the kets, we get
\ben
\vert\Psi\rangle_{out} &=& [\alpha\alpha^\prime (r^2_v  - t^2_v) \vert a\rangle^V \vert b\rangle^V + \beta\beta^\prime (r^2_h - t^2_h) \vert a\rangle^H \vert b\rangle^H]\nonumber\\&+& [(\alpha\beta^\prime r_v r_h - \beta\alpha^\prime t_h t_v) \vert a\rangle^V \vert b\rangle^H - (\alpha\beta^\prime t_v t_h - \beta\alpha^\prime r_h r_v)\vert a\rangle^H \vert b\rangle^V)]\nonumber\\ &+& i(\alpha\beta^\prime r_v t_h + \beta\alpha^\prime r_h t_v) \vert a\rangle^V \vert a\rangle^H + i(\alpha\beta^\prime r_h t_v + \beta\alpha^\prime r_v t_h) \vert b\rangle^V \vert b\rangle^H\nonumber\\ &+& i\alpha\alpha^\prime r_v t_v [\vert a\rangle^V \vert a\rangle^V + \vert b\rangle^V \vert b\rangle^V]\nonumber\\ &+& i\beta\beta^\prime r_h t_h [\vert a\rangle^H \vert a\rangle^H + \vert b\rangle^H \vert b\rangle^H].\label{f} 
\een
Only the first two terms in square braces represent entangled states, the rest being bunched photon terms representing both photons on the same side of the beam splitter. The state (\ref{f}) can be written as
\beq
\vert\Psi\rangle_{out} = c_{\Phi^+} \vert \Phi^+\rangle + c_{\Phi^-} \vert \Phi^-\rangle + c_{\Psi^+} \vert \Psi^+\rangle + c_{\Psi^-} \vert \Psi^-\rangle + \cdots \label{bell}
\eeq
where, juxtaposing the notations for the path and polarization appropriately,
\ben
\vert \Phi^+\rangle &=& \frac{1}{\sqrt{2}}[\vert a\rangle^H\vert b\rangle^H + \vert a\rangle^V\vert b\rangle^V]= \frac{1}{\sqrt{2}}[\vert H\rangle^a\vert H\rangle^b + \vert V\rangle^a\vert V\rangle^b],\\ 
\vert \Phi^-\rangle &=& \frac{1}{\sqrt{2}}[\vert a\rangle^H\vert b\rangle^H - \vert a\rangle^V\vert b\rangle^V] = \frac{1}{\sqrt{2}}[\vert H\rangle^a\vert H\rangle^b - \vert V\rangle^a\vert V\rangle^b],\\ 
\vert \Psi^+\rangle &=& \frac{1}{\sqrt{2}}[\vert a\rangle^V\vert b\rangle^H + \vert a\rangle^H\vert b\rangle^V] = \frac{1}{\sqrt{2}}[\vert V\rangle^a\vert H\rangle^b + \vert H\rangle^a\vert V\rangle^b],\\ 
\vert \Psi^-\rangle &=& \frac{1}{\sqrt{2}}[\vert a\rangle^V\vert b\rangle^H - \vert a\rangle^H\vert b\rangle^V] = \frac{1}{\sqrt{2}}[\vert V\rangle^a\vert H\rangle^b - \vert H\rangle^a\vert N\rangle^b]\label{f2}
\een
are the Bell states, the dots represent the bunched photon states, and
\ben
c_{\Phi^+} &=& \frac{1}{\sqrt{2}}[\alpha\alpha^\prime (r^2_v - t^2_v)  + \beta\beta^\prime (r^2_h  - t^2_h)],\label{d1}\\
c_{\Phi^-} &=& \frac{1}{\sqrt{2}}[\beta\beta^\prime (r^2_h  - t^2_h) -\alpha\alpha^\prime (r^2_v - t^2_v)],\label{d2}\\
c_{\Psi^+} &=& \frac{(\alpha\beta^\prime + \beta\alpha^\prime)}{\sqrt{2}}[ r_v r_h - t_v t_h],\label{d3}\\
c_{\Psi^-} &=& \frac{(\alpha\beta^\prime - \beta\alpha^\prime)}{\sqrt{2}}[r_v r_h + t_v t_h].\label{d4} 
\een
Note that the four Bell states occur with different coefficients, and hence entanglement is not entirely cancelled.
Note also that even with a fixed initial polarization state, i.e. $\alpha = \alpha^\prime = 1, \beta = \beta^\prime = 0$ or $\beta = \beta^\prime = 1, \alpha = \alpha^\prime = 0$, a state like $\vert V\rangle^a \vert V\rangle^b$ or $\vert H\rangle^a \vert H\rangle^b$ is produced and the photon bunching rule \cite{bunching, hom} does not hold. This is because the  special relationship responsible for photon bunching does not hold for a non-symmetric beam splitter. This is easily checked by using the conditions for a symmetric 50-50 beam splitter, namely $r_v = t_v$ and $ r_h = t_h$ in Eqn. (\ref{f}) and verifying that these terms do indeed disappear. On the other hand, even with a symmetric 50-50 beam splitter, the state $\vert \Psi^-\rangle$ can still be produced unless $\alpha\beta^\prime = \beta\alpha^\prime$. {\em The use of a non-symmetric beam splitter together with the use of incoming states of arbitrary polarization that can be manipulated turns out to be critical}.

We will now consider two alternative arrangements.
\begin{enumerate}
\item
Set $\alpha = \beta = \alpha^\prime = \beta^\prime = 1/\sqrt{2}$ so that $\alpha \alpha^\prime = \beta\beta^\prime = 1/2$ and $\alpha\beta^\prime = \beta\alpha^\prime = 1/2$. Also, choose a beam splitter and the angles of incidence such that $r^2_v - t^2_v = r^2_h - t^2_h \neq 0 $. Then, $c_{\Psi^-} = c_{\Phi^-} = 0$ but $c_{\Phi^+} \neq 0, c_{\Psi^+} \neq 0$.  
\item
Set $\alpha = \beta = \alpha^\prime = -\beta^\prime = 1/\sqrt{2}$ so that $\alpha \alpha^\prime = -\beta\beta^\prime = 1/2$ and $\alpha\beta^\prime = - \beta\alpha^\prime = - 1/2$ and also choose $r^2_v - t^2_v = r^2_h - t^2_h \neq 0$ as before. Then, $c_{\Phi^+} = c_{\Psi^+} = 0$ but $c_{\Phi^-} \neq 0, c_{\Psi^-} \neq 0$.
\end{enumerate}
The Bell states can now be discriminated in the coincidence basis as follows. Let the outgoing modes from $BS$ pass through two 50-50 polarizing beam splitters $PBS_a$ and $PBS_b$ which separate out the horizontal and vertical polarization states (Fig. 1). Let $A_h, A_v, B_h, B_v$ be single-photon detectors. Then, in case 1,
coincidences at $A_hB_v$ and $A_vB_h$ signal $\vert \Psi^+\rangle$ and coincidences at $A_hB_h$ and $A_vB_v$ signal $\vert \Phi^+\rangle$. In case 2, coincidences at $A_hB_v$ and $A_vB_h$ signal $\vert \Psi^-\rangle$ and coincidences at $A_hB_h$ and $A_vB_v$ signal $\vert \Phi^-\rangle$. The bunched photon states produce double counts and coincidences at $A_vA_h$ and $B_vB_h$ and can be ignored. 

Finally, consider the fidelity of the target states $\vert \Phi^+\rangle$ and $\vert \Psi^+\rangle$ in case 1 as the parameters $\alpha, \beta, \alpha^\prime, \beta^\prime$ are varied around $1/\sqrt{2}$. Using the reparametrizations $\alpha = \alpha^\prime = \frac{1}{\sqrt{2}}={\rm cos}\frac{\pi}{4}, \beta = \beta^\prime = \frac{1}{\sqrt{2}} = {\rm sin}\frac{\pi}{4}$, we can rewrite (\ref{f}) as   
\ben
\vert\Psi\rangle &=& c_{\Phi^+}\vert\Phi^+\rangle +  c_{\Psi^+}\vert\Psi^+\rangle + \cdots\nonumber\\
&=& \left(\frac{r^2_{v}-t^2_{v}}{\sqrt{2}}\right)\vert\Phi^+\rangle +
 \left(\frac{r_{v}r_{h}-t_{v}t_{h}}{\sqrt{2}}\right)\vert\Psi^+\rangle + \cdots
\een
Varying the parameters, we get 
\ben
\vert \Psi\rangle^\prime &=& (r^2_{v}-t^2_{v}).\nonumber\\
&&[\cos(\frac{\pi}{4}+\epsilon)\cos(\frac{\pi}{4}+\epsilon^{'})\vert V\rangle^{a}\vert V\rangle^{b} + \sin(\frac{\pi}{4}+\epsilon)\sin(\frac{\pi}{4}+\epsilon^{'})\vert H\rangle^{a}\vert H\rangle^{b}] \nonumber\\
&+& [(\cos(\frac{\pi}{4}+\epsilon)\sin(\frac{\pi}{4}+\epsilon^{'})r_{v}r_{h}-\sin(\frac{\pi}{4}+\epsilon)\cos(\frac{\pi}{4}+\epsilon^{'})t_{v}t_{h})\vert V\rangle^{a}\vert H\rangle^{b}] \nonumber\\
 &-& [(\cos(\frac{\pi}{4}+\epsilon)\sin(\frac{\pi}{4}+\epsilon^{'})t_{v}t_{h}-\sin(\frac{\pi}{4}+\epsilon)\cos(\frac{\pi}{4}+\epsilon^{'})r_{v}r_{h})\vert H\rangle^{a}\vert V\rangle^{b}]\nonumber\\
&+& \cdots
\een
Therefore,  
\beq
\frac{\langle\Phi^+\vert\Psi\rangle^\prime}{\langle\Phi^+\vert\Psi\rangle} = \cos(\epsilon-\epsilon^{'})
\eeq
and
\beq
\frac{\langle\Psi^+\vert\Psi\rangle^\prime}{\langle\Psi^+\vert\Psi\rangle} = \cos(\epsilon+\epsilon^{'}).
\eeq
Similarly, for case 2 in which $\vert\Phi^-\rangle$ and $\vert\Psi^-\rangle$ are produced, we have
\beq
\frac{\langle\Phi^-\vert\Psi\rangle^\prime}{\langle\Phi^-\vert\Psi\rangle} = \cos(\epsilon+\epsilon^{'})
\eeq
and
\beq
\frac{\langle\Psi^-\vert\Psi\rangle^\prime}{\langle\Psi^-\vert\Psi\rangle} = \cos(\epsilon-\epsilon^{'}).
\eeq 
Hence, the fidelity of these states is $\cos^2 (\epsilon \pm \epsilon^{'})$.
  
This completes the theoretical demonstration. 
\section{Conclusion}
The new result is that arbitrary superpositions of the Bell states can be generated from consective and indistinguishable single-photon pulses whose polarization states can be manipulated by making them overlap on a nonpolarizing and non-symmetric beam splitter $BS$ from opposite sides (Fig. 1). This is impossible with symmetric beam splitters. By using a beam splitter $BS$ with the right reflection and transmission coefficients and by also tuning the amplitudes of the $H$ and $V$ components of the incident photons, a superposition of two Bell states with different amplitudes can be produced for each arrangement. These two Bell states of high fidelity can be discriminated in the coincidence basis by post-selection. This method therefore has the potential for use as a versatile resource for quantum information processing without the need for nonlinear processes such as SPDC at any stage, and also for testing the foundations of quantum mechanics. 

\section{Acknowledgement}
I thank the National Academy of Sciences, India for the award of a Senior Scientist Platinum Jubilee Fellowship which allowed this work to be undertaken. I also thank Anirban Mukherjee for help with the fidelity calculation.


\begin{thebibliography}{0}
\bibitem{qc1}
C. Kurtsiefer et al (2002), {\em Quantum cryptography: A step
towards global key distribution}, Nature, 419, 450.
\bibitem{qc2}
M. Aspelmeyer et al (2003), {\em Long-distance free-space distribution of
quantum entanglement}, Science, 301, pp. 621-623.
\bibitem{qc3}
I. Marcikic et al (2004), {\em Distribution of time-bin entangled qubits over 50 km of optical fiber} Phys. Rev. Lett., 93, 180502-(1-4).

\bibitem{pdc1}
D. C. Burnham  \& D. L. Weinberg (1970), {\em Observation of simultaneity in parametric production of optical photon
pairs}, Phys. Rev. Lett., 25, pp. 84-87.
\bibitem{pdc2}
S. Friberg, C. K. Hong, \& L. Mandel (1985), {em Measurement of time delays in the parametric production of photon
pairs}, Phys. Rev. Lett., 54, pp. 2011-2013.
\bibitem{pdc3}
S. Friberg \& L. Mandel (1984), {\em Production of squeezed states by combination of parametric down-conversion and
harmonic generation}, Opt. Commun., 48, pp. 439-442.
\bibitem{pdc4}
P. G. Kwiat, E, Waks, A. G. White, I. Appelbaum, \& P. B. Eberhard (1999), {\em Ultrabright source of polarization entangled photons}, Phys. Rev. A, 60, R773.
\bibitem{pdc5}
C. Kurtsiefer, M. Oberparleiter, \& H. Weinfurter (2001), {\em High-efficiency entangled photon pair collection in type-II parametric fluorescenc}, Phys. Rev. A, 64, 023802.
\bibitem{pdc6}
S. Tanzilli et al (2001), {\em Highly efficient photon-pair source using periodically poled lithium niobate waveguide}, Elec. Lett., 37, pp. 26-28.
\bibitem{pdc7}
S. J. Mason, N. A. Albota, F. Konig, \& F. N. C. Wong (2002),  {\em Efficient generation of tunable photon pairs at 0.8 and 1.6 ìm}, Opt. Lett., 27, 2115.
\bibitem{pdc8}
F. K\"onig, E. J. Mason, F. N. C. Wong, \& M. A. Albota (2005), {\em Efficient and spectrally bright source of polarization-entangled photons}, Phys. Rev. A, 71, 033805.

\bibitem{2photon1}
E. Brannen, F. R. Hunt, R. H. Adlington, \& R. W. Hicholls (1955), {\em Application of Nuclear coincidence methods to
atomic transitions in the wavelength range 2000-6000A}, Nature, 175, 810.
\bibitem{2photon2}
A. Kuzmich et al (2003), {\em Generation of nonclassical photon pairs for scalable quantum communication with atomic ensembles}, Nature, 423, pp. 731-734.

\bibitem{fan}
J. Fan, A. Migdall, \& L. J. Wang (2005), {\em A microstructure fiber two photon source with conjugate laser pumps},  Quantum Communications and Quantum Imaging III, edited by Ronald E. Meyers, Yanhua Shih, Proc. Of SPIE, 5893, 589309. 

\bibitem{chen}
J. Chen, K. F. Lee, C. Liang, \& P. Kumar (2006), {\em Fiber-based telecom-band degenerate-frequency source of entangled photon pairs}, Opt. Lett., 31, pp. 2798-2800.

\bibitem{qdots}
C. Santori, D. Fattal, J. Vu\u{c}kovi\'{c}, G. S. Solomon, \& Y. Yamamoto (2002), {\em Indistinguishable photons from a single-photon device}, Nature, 419, pp. 594-597.

\bibitem{zeilinger}
A. Zeilinger (1981), {\em General properties of lossless beam splitters in interferometry}, Am. J. Phys. 49 (9), pp. 882-883.\\
C. H. Holbrow, E. Galvez \& M. E. Parks (2001), Am. J. Phys. 70 (3), pp. 260-265. 

\bibitem{yurke}
B. Yurke and D. Stoler (1992), {\em Einstein-Podolsky-Rosen effects from independent particle sources}, Phys. Rev. Lett. 68, pp. 1251-1254; {\em Bell's-inequality experiments using independent-particle sources}, Phys. Rev. A 46, pp. 2229-2234.

\bibitem{knill}
E. Knill, R. Laflamme \& G. J. Milburn (2001). {\em A scheme for efficient quantum computation with linear optics}, Nature, 409, pp. 46-52.

\bibitem{bunching}
G. Weihs, M. Reck, H. Weinfurter, \& A. Zeilinger (1996), {\em Two-photon interference in optical fiber multiports}, Phys. Rev. Lett., 54, pp. 893-897.

\bibitem{hom}
C. K. Hong, Z. Y. Ou, \& L. Mandel (1987), {\em Measurement of subpicosecond time intervals between two
photons by interference}, Phys. Rev. Lett., 59, pp. 2044–2046.

\end{thebibliography}
\end{document}